\documentclass[namedreferences]{solarphysics}
\usepackage[optionalrh]{spr-sola-addons} 
\usepackage{graphicx}        
\usepackage{color}           

\newcommand{\figfolder}{}

\newcommand{\aap}{    {\it Astron. Astrophys.}}

\newcommand{\apj}{    {\it Astrophys. J.}}
\newcommand{\apjl}{   {\it Astrophys. J. Lett.}}

\newcommand{\solphys}{{\it Solar Phys.}}

\newcommand{\ssr}{    {\it Space Sci. Rev.}}
\chardef\us=`\_

\begin{document}

\begin{article}
\begin{opening}

\title{An Optimization Principle for Computing Stationary MHD
Equilibria With Solar Wind Flow}

\author[addressref={aff1},corref,email={wiegelmann@mps.mpg.de}]
{\inits{}\fnm{Thomas}~\lnm{Wiegelmann}\orcid{0000-0001-6238-0721}}
\author[addressref=aff2,email={}]
{\inits{}\fnm{Thomas}~\lnm{Neukirch}\orcid{0000-0002-7597-4980}}
\author[addressref=aff3,email={}]
{\inits{H.}\fnm{Dieter H.}~\lnm{Nickeler}\orcid{0000-0001-5165-6331}}
\author{Iulia~\surname{Chifu}$^{1,4}$\orcid{0000-0002-5481-9228}}
\address[id=aff1]{Max-Planck-Institut f\"ur Sonnensystemforschung,
Justus-von-Liebig-Weg 3,
37077 G\"ottingen, Germany.}
\address[id=aff2]{School of Mathematics and Statistics,
University of St. Andrews, St. Andrews, KY16 9SS, UK}
\address[id=aff3]{Astronomical Institute, Czech Academy of Sciences,
Fri\v{c}ova 298, 25165 Ond\v{r}ejov, Czech Republic}
\institute{$^{4}$ Astronomical Institute of Romanian Academy,
Cutitul de Argint 5, Bucharest, Romania}
\runningauthor{T. Wiegelmann et al.}
\runningtitle{Optimization Code for Stationary MHD Equilibria
With Solar Wind Flow}
\begin{abstract}
In this work we describe a numerical optimization method for
computing stationary MHD-equilibria. The newly developed code is
based on a nonlinear force-free optimization principle.
We apply our code to model the solar corona using synoptic vector
magnetograms as boundary condition. Below about two solar radii the
plasma $\beta$ and Alfv\'en Mach number $M_A$ are small and the
magnetic field configuration of stationary MHD is basically identical to
a nonlinear force-free field, whereas higher up in the corona
(where $\beta$ and $M_A$ are above unity) plasma
and flow effects become important and stationary MHD and force-free
configuration deviate significantly.
The new method allows the reconstruction of the coronal magnetic field further
outwards than with potential field, nonlinear force-free or
magneto-static models.
This way the model might help to provide the magnetic connectivity
for joint observations of remote sensing and in-situ instruments on Solar
Orbiter and Parker Solar Probe.
\end{abstract}
\end{opening}

\section{Introduction}
Traditionally the global structure of the coronal magnetic field
is modelled with the help of source surface potential field models

\citep[PFSS, see][]{1969SoPh....6..442S}. In potential field models electric
currents are neglected and they require as photospheric boundary
condition only line-of-sight magnetic field measurements.
Nevertheless, the effect of a solar wind
flow is considered by the introduction of a so called
`source surface'
as outer boundary condition.
At this artificial
surface, usually at $2.5 R_s$, all field lines are assumed
to become radial, thereby
mimicking the effect of the solar wind.
Going beyond this simple potential field approach with non-potential
global coronal magnetic field models has been an active research topic for
several years with various physical models
\citep[see, e.g.,][for review articles on global coronal magnetic field
modelling.]{2012LRSP....9....6M,2017SSRv..210..249W}.
Recently several of the non-potential methods
(seven different codes) have been compared
in the framework of an ISSI-meeting and the results have been
published \citep[][]{2018SSRv..214...99Y}.
While all of the models are non-potential and incorporate electric currents,
the physical assumptions and computational
implementations are different and
the models used also different input data. One group of codes are
nonlinear force-free extrapolations (3 different implementations have
been compared), which do not consider time dependence and
plasma effects. As boundary condition these codes require measurements of
the photospheric magnetic field vector. Codes for solving the nonlinear
force-free equations have been first applied to active regions and
different numerical methods and implementations have been intensively
compared and evaluated in a number of studies, e.g.,
\cite{2006SoPh..235..161S,2008SoPh..247..269M,2008ApJ...675.1637S,
2009ApJ...696.1780D,2015ApJ...811..107D}. Active region and
global nonlinear force-free
codes finally solve the same nonlinear force-free equations, but
do so with different numerical implementations. Well known codes
for global nonlinear force-free computations are based on
the Grad-Rubin method \citep[see][]{2013A&A...553A..43A,2014JPhCS.544a2012A},
an optimization principle \citep[see][]{2007SoPh..240..227W,2014A&A...562A.105T}
and force-free electrodynamics
\citep[see][]{2011SoPh..269..351C,2013SoPh..282..419C}.

Other approaches go beyond the force-free assumption and take effects
like plasma forces, flows and time-dependence into account.
In the linear magnetostatic approaches
\citep[see][]{bogdan:etal86,neukirch95} the
Lorentz force is compensated by plasma pressure and gravity forces.
Coronal equilibria with plasma
pressure and steady 3D nonlinear flows have been found in
\cite{2017ApJ...837..104N}.
The evolving magnetofrictional method
\citep[see][]{2006ApJ...641..577M,2014SoPh..289..631Y} solves also the force-free
equations, but takes time-dependence into account and requires therefore
a time sequence of photospheric magnetograms (radial component only)
as boundary condition and incorporates flux transport. MHD codes
from different research groups
(see, e.g., \cite{1994ApJ...430..898M,1999PhPl....6.2217M}
and \cite{2012SoPh..279..207F,2020Feng_book}) are capable to
solve the full MHD equations.

For global corona models these codes use the radial photospheric field as
boundary conditions. It is possible, however, to limit the MHD simulations
to the assumption of a zero-$\beta$ plasma, and to incorporate additional
observations, e.g., use results from magnetofrictional simulations
\citep[see][for details]{2018SSRv..214...99Y}.

An interesting point is that only the (static) nonlinear force-free codes
make use of the photospheric vector magnetograms, whereas
all other methods compared in \cite{2018SSRv..214...99Y}
use the radial photospheric field only.
As a consequence the
study of \cite{2018SSRv..214...99Y} revealed that the nonlinear force-free
extrapolations are superior in active regions (where the vector magnetic
field measurements are most accurate) while quiet Sun features like
filament channels are better modelled by other approaches. These findings
stimulated us using vector magnetograms as boundary condition and
go beyond the force-free assumption. To do so we develop a new stationary
MHD code with field aligned plasma flow, which uses synoptic vector
magnetograms as boundary condition.

We organize the paper as follows:
In Section \ref{basics} we present the stationary MHD-equations and
how we aim to solve them with the help of an optimization principle.
Section \ref{test} contains an application  and evaluation of the newly
developed code to a synoptic vector magnetogram observed with SDO/HMI.
Finally we draw conclusions and give an outlook for future work in
Section \ref{conclusions}.

\section{Basic Equations}
\label{basics}
\subsection{Stationary Compressible MHD}
To model the coronal magnetic field and plasma environment
we use the equations of stationary compressible ideal MHD.
 \begin{eqnarray}
\rho\left({\bf v} \cdot \nabla\right){\bf v} &=&
\frac{1}{\mu_0} \, \left(\nabla\times {\bf B} \right)\times {\bf B}
- \nabla p -\rho\nabla\Psi\ ,\label{force} \\
\nabla\cdot {\bf B} &=& 0 \label{divB} \, , \\
\nabla\cdot\left(\rho{\bf v}\right)&=
& 0\, , \label{conti} \\
p & = & \rho R \, T, \label{energy} \\
{\bf E} +  {\bf v} \times {\bf B} &=& 0.
\label{ohm}
\end{eqnarray}
where ${\bf B}$ is the magnetic field, $p$ the plasma pressure,
$\rho$ the mass density, ${\bf v}$ the flow velocity,
$\Psi$ the gravity potential, $\mu_0$ the permeability of free space,
$T$ the temperature, $R$ the gas constant and ${\bf E}$ the electric
field. The stationary MHD equations are given by
 the force balance Equation \ref{force}, the solenoidal condition
 (Equation \ref{divB}), mass continuity Equation \ref{conti}, an energy
 equation or an equation of state (Equation \ref{energy}) and ideal Ohm's law
 (Equation \ref{ohm}).
 We are aware that non-ideal effects as, e.g., caused by turbulence,
 can lead to a violation of ideal Ohm's law and play an important role
 for coronal heating and solar wind expansion
 \citep[see][for a review article]{2015RSPTA.37340148C}.
 For special cases ($2.5D$, incompressible flow and no gravity) stationary
solutions of resistive MHD have been found
\citep[see, e.g.,][]{1998JPlPh..59..303T,2000JPlPh..64..601T,2014A&A...569A..44N}.
Studying turbulence and resistivity is
well outside the scope of this work, however.
For simplicity we assume an isothermal
equation of state in
Equation \ref{energy}, which leads to a linear relation of plasma
pressure and density.
We replace $p$ by $\rho R T$ in
Equation \ref{force} to reduce the number of independent quantities.
For the highly conducting coronal plasma we have
to consider ideal Ohms law, which is for a vanishing electric field
satisfied if
${\bf v} \times {\bf B} = \nabla f$,
with an arbitrary scalar function $f$, which is constant on magnetic
field lines. For the particular choice $f=0$, which we use here,
the plasma flows and magnetic fields are parallel.
\subsection{Optimization Principle for Stationary MHD}
Minimizing a functional $L$ of quadratic terms to compute 3D coronal
magnetic field models has been introduced by \cite{2000ApJ...540.1150W}
for computing nonlinear force-free fields. First implementations of
this optimization approach have been done in Cartesian geometry to
model active regions.
A spherical implementation for global force-free
optimization has been done in \cite{2007SoPh..240..227W} and was first
applied with synoptic vector magnetograms as boundary condition in
\cite{2014A&A...562A.105T}. A magneto-hydro-static optimization code
has been developed in \cite{2007A&A...475..701W} for global computations
in spherical geometry. Within this work we try to extend this
optimization approach towards stationary MHD.

We aim to solve the stationary MHD Equations
\ref{force}-\ref{ohm} by minimizing a functional $L$
which we define as
\begin{equation}
L({\bf B}, {\bf v}, \rho) =
L_{\rm force} + L_{\rm divB} + L_{\rm cont} + L_{\rm angle(B, v)},
\label{defL}
\end{equation}
where all terms in the functional have a quadratic form as defined below.
This means that the stationary MHD equations are solved when $L$ is zero.
\small
\begin{equation}
L_{\rm force} = \int_{V} \frac{ \left[ \left({\nabla} \times {\bf B}\right)\times {\bf B}
- \mu_0 \nabla (\rho R T) - \mu_0 \rho \nabla\Psi
-\mu_0 \rho \left(\nabla \times {\bf v} \right) \times {\bf v}
-\frac{\mu_0 \rho}{2} \nabla {v^2} \right]^2}{B^2} \; dV
\label{L_force}
\end{equation}
\normalsize
If the term $L_{\rm force}$ vanishes, Equation \ref{force} is satisfied.
We applied vector
identities to the flow terms to bring them
into a similar form as the
magnetic terms. The linearity in the
isothermal equation of state
(Equation \ref{energy}) has been used to substitute $p$ by $\rho R T$.

The other parts of the functional are linked to the solenoidal condition,
the steady-state continuity equation, and the condition that the flow
is field aligned:
\begin{eqnarray}
L_{\rm divB} & = & \int_{V} [ \nabla \cdot {\bf B}]^2 \; dV,
\label{L_divB} \\
L_{\rm cont} & = &
\int_{V} \left[\nabla\cdot\left(\rho {\bf v} \right) \right]^2 \; dV,
\label{L_cont} \\
L_{\rm angle(B, v)} & = & \int_{V} \; \tanh(M_A^2) \;
\frac{\left[\bf v \times \bf B \right]^2}{v^2 \, B^2} \; dV,
\label{L_angle}
\end{eqnarray}
where $M_A=v/v_A$ is the Alfv\'en Mach number and
$v_A=B/\sqrt{\mu_0 \rho}$ is the Alfv\'en velocity.
In the $L_{\rm angle(B, v)}$ term we use a weighting with the
Alfv\'en Mach number $M_A$ to give a stronger weight to regions
with strong flow.
The $B^2$ in the nominator of Equation \ref{L_force}
originates from the nonlinear force-free and magneto-hydro-static
optimization codes. Dividing by this term ensures that sufficient weight is given
to the equilibrium in weak field regions and gives
the terms $L_{\rm force}$ and $L_{\rm divB}$ the same dimensionality.
While in the functional all terms are of quadratic form, it might be
convenient for humans to monitor also the angle between magnetic
fields and flows in degree.
\begin{equation}
{\rm angle(B, v)}= {\rm asin} \left(\left(\int \frac{|{\bf v} \times {\bf B}|}{|{\bf B}|} \; dV \right)/
\int |{\bf v}| \; dV \right)
\label{defangle}
\end{equation}
This formula is similar
to the definition of the weighted average
angle between magnetic field and electric currents used for evaluating
the quality of force-free computations
\citep[see][for details.]{2006SoPh..235..161S}.

\section{Testing and Application of the Method}
\label{test}
\begin{figure}[h]
\includegraphics[width=12cm, height=5cm, bb = 20 150 980 500, clip]{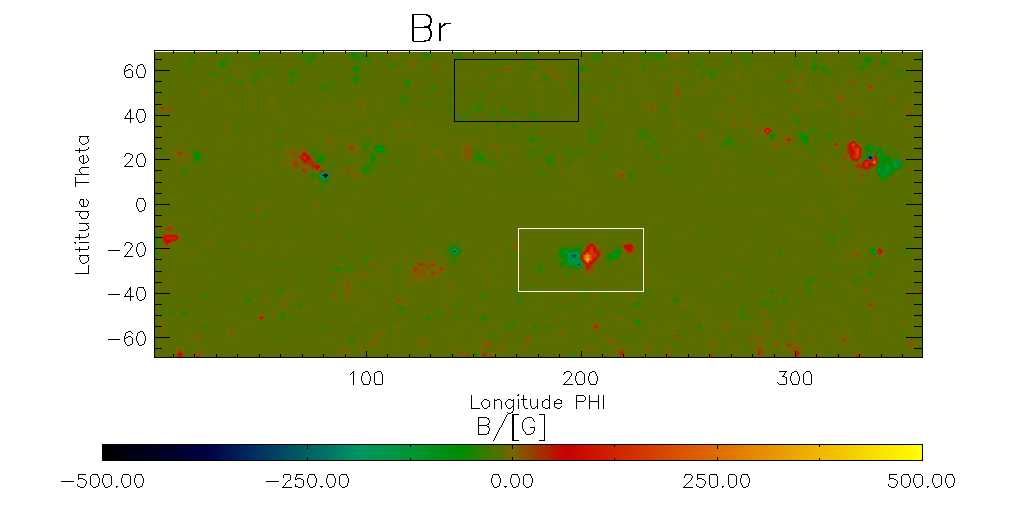}
\includegraphics[width=12cm, height=5cm, bb = 20 150 980 500, clip]{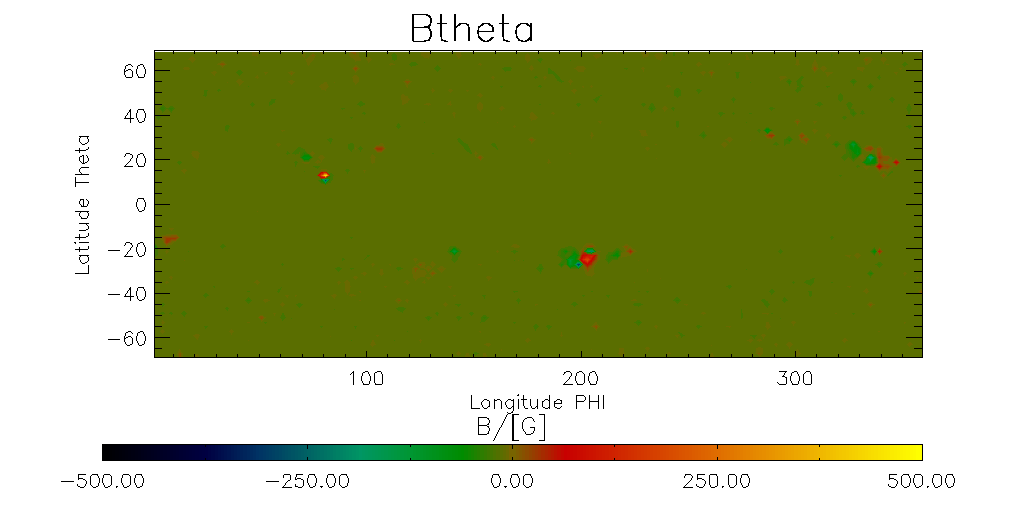}
\includegraphics[width=12cm, height=7.5cm, bb = 20 0 980 500, clip]{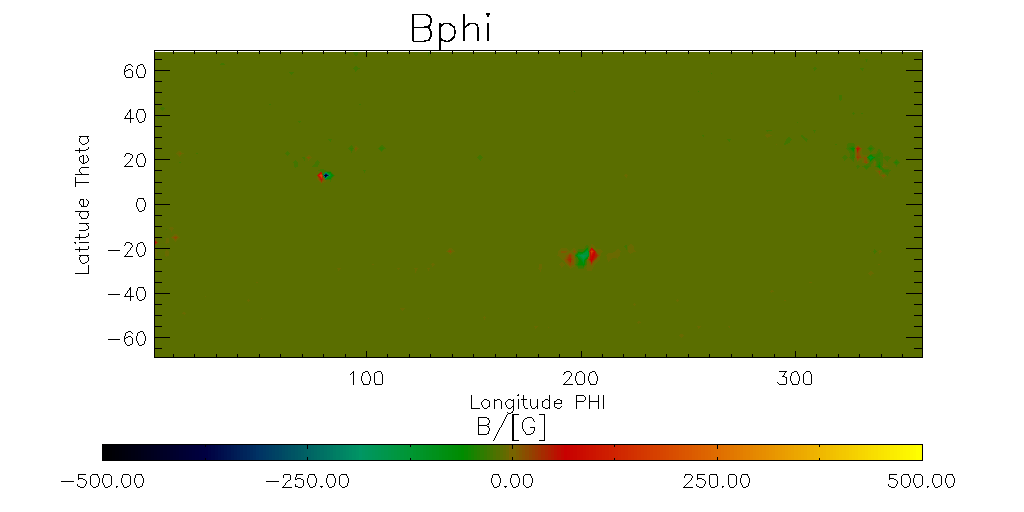}
\caption{As boundary data we use a synoptic vector magnetogram
 from SDO/HMI for Carrington rotation 2099, which has been observed
between 13/07/2010 and 09/08/2010. The polar regions have been cut out.
One active region (white box in top panel) and a quiet sun area of the
same size (black box) are investigated in Section \ref{sec_rev}.}
\label{fig1}
\end{figure}
\begin{figure}[h]
\mbox{
\includegraphics[width=0.45 \textwidth]{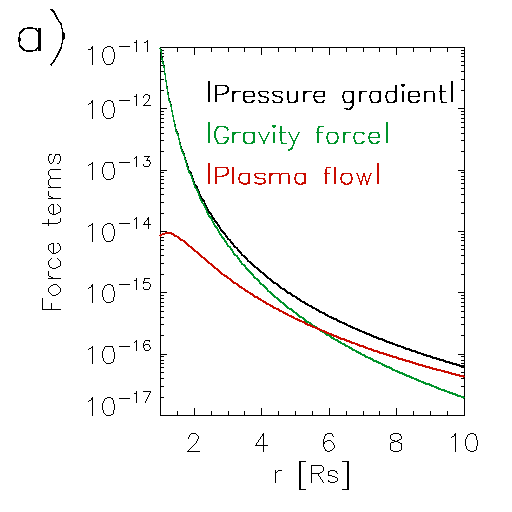}
\includegraphics[width=0.45 \textwidth]{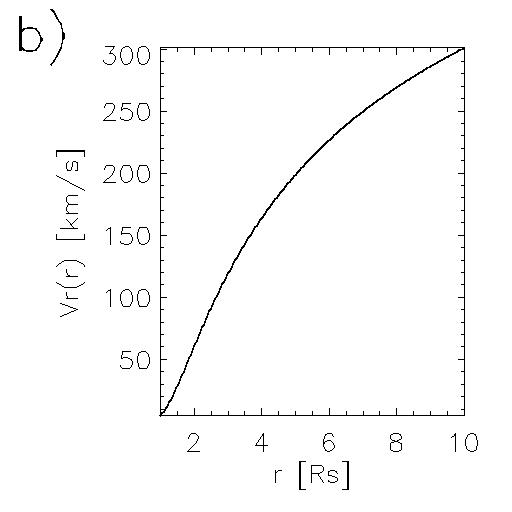}}
\mbox{
\includegraphics[width=0.45 \textwidth]{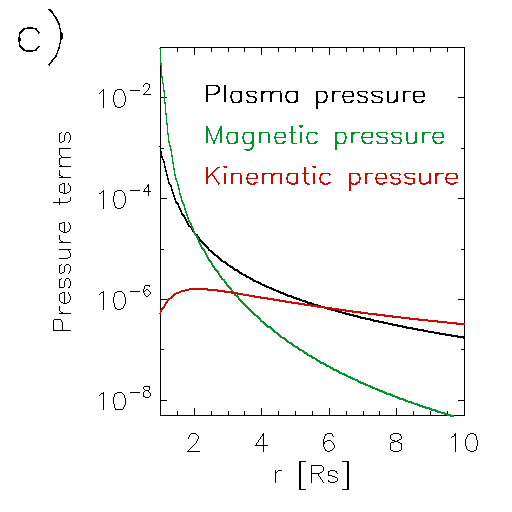}
\includegraphics[width=0.45 \textwidth]{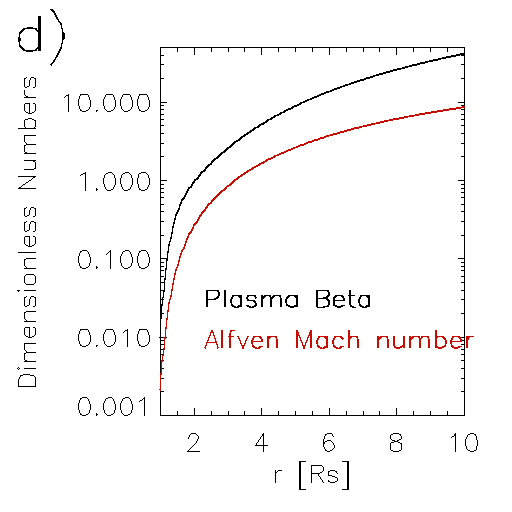}}
\caption{
The variation of the initial conditions, based on a combination of a
force-free magnetic field and
a spherically symmetric Parker wind solution.
a) Shows the different force terms
b) Solar wind velocity as function of r
c) Plasma and kinematic pressure from Parker's model,
magnetic pressure from force-free field model.
d) Plasma Beta and Alfv\'en Mach number:
If both quantities are small, plasma and flow forces
can be neglected for computing the coronal magnetic field.
}
\label{fig2}
\end{figure}
Previous versions of optimization codes have been tested first with
known analytical or semi-analytical equilibria, e.g.
\cite{1990ApJ...352..343L}
for nonlinear force-free active region models and
\cite{neukirch95} for global
magnetostatic equilibria.
Unfortunately we are not aware of analytic 3D solutions
of stationary MHD-equilibria with compressible plasma
flow and gravity. We therefore use the code to construct
a numerical equilibrium by minimizing the functional
(Equation \ref{defL}) and monitoring the individual terms
(Equations \ref{L_force} - \ref{L_angle}).

\subsection{Boundary Conditions and Initial Force-Free Model}
If all non-magnetic terms in the stationary MHD Equations
\ref{force}-\ref{ohm} are neglected, we get as a subclass
the nonlinear force-free field equations, which are given as
\begin{eqnarray}
\left(\nabla\times\bf B\right)\times\bf B & = & 0 \label{nlfff1} \\
\nabla\cdot\bf B &=& 0. \label{nlfff2}
\end{eqnarray}
 As boundary data we use a synoptic vector magnetogram
 from SDO/HMI for Carrington rotation 2099 as shown in Figure \ref{fig1},
 which has been observed between 13/07/2010 and 09/08/2010. Because of the
 grid convergence problem at the poles
 (see \cite{2007SoPh..240..227W}) and because accurate vector field
 measurements at poles are still not available, we cut out polar regions
 and limit our computation to latitudes $\theta=20 \dots 160$. The
 spatial grid resolution is 2 degrees.
 As initial state we compute
 a nonlinear force-free field up to $r=10 R_s$, which solves the force-free
 Equations \ref{nlfff1} - \ref{nlfff2} with the global force-free
  optimization approach as described in
 \citep[][]{2007SoPh..240..227W,2014A&A...562A.105T}.
The force-free iteration itself uses a potential field as initial state and
the potential field solution is also kept on the theta boundaries.
We would like to point out that the initial state for the force-free iteration
 is not a PFSS model, because that is radially limited to the source surface
 located at $2.5 R_s$. We use just the decaying mode of a spherical harmonic
 representation and for the tests done here limit them to $l=12$. This
 corresponds to a special case of the global linear magnetostatic model
 developed in \cite{bogdan:etal86} with $\alpha=0$ and $a=0$.

\subsection{Initial Parker Solar Wind Solution}
If all magnetic terms in the stationary MHD Equations
\ref{force}-\ref{ohm} are neglected, we get as a subclass
the stationary hydrodynamic  equations
\begin{eqnarray}
\rho\left(\bf v\cdot\bf\nabla\right)\bf v &=&
- \nabla p -\rho\nabla\Psi \label{parker1} \\
\nabla\cdot\left(\rho\bf v\right)&= & 0 \label{parker2}\\
p & = & \rho R \, T, \label{parker3}
\end{eqnarray}
where the gravitational potential is
$\Psi=-G M_s/r$, where $G$ is the gravity constant, $M_S$ the solar
mass and $r$ the distance from the center of the sun.
To initialize the plasma and flow variables we use the spherical symmetric
solution of the stationary hydrodynamic Equations
\ref{parker1} - \ref{parker3}, which was found by
\cite{1958ApJ...128..664P} and describes the solar wind.
We use an isothermal solution
which can be solved analytically by using the Lambert W function
\citep[see][for details]{2004AmJPh..72.1397C}.
In the isothermal case there are only two free parameters,
$T$ and $p$ (or alternatively $T$ and $\rho$).
We use $T=3$MK and $p$ is so specified that the average plasma beta
at $r= 1 R_s$ is $\beta=0.01$, whereas the magnetic pressure was computed
(and averaged over the sphere)
from the initial force-free magnetic field.
The sound velocity is $c_s=157$ km/s and the corresponding critical
radius is at $r_c=3.84 R_s$. Figure \ref{fig2} shows several quantities
of this spherically
symmetric solution as a function of the radius $r$.
Panel a) contains the plasma pressure gradient force (black), gravity
force (green) and the flow force (red). At low coronal heights the flow
force is small and gravity and pressure forces compensate each other.
With increasing distance from the sun
 the flow becomes more and more important. The flow
velocity is shown in panel b). In panel c) we compare the
magnetic pressure (green), plasma pressure (black) and solar wind ram
pressure (red). At low heights in the low $\beta$ corona (black line in
panel d) the magnetic pressure dominates. This is also the region which
is usually computed using the force-free assumption. With increasing height
the plasma pressure and kinematic pressure become more important and
finally dominate over the magnetic pressure. The red line in panel d)
shows the Alfv{\'e}n Mach-number $M_A$. In the lower corona
$M_A$ is very small, but increases to values above unity with increasing
distance from the sun. In regions with $\beta \ll 1$ and $M_A \ll 1$
it is justified to neglect the influence of plasma effects and flows
and this is a justification why these regions are usually modelled under
the force-free assumption. Higher up in the corona, however,
$\beta$ and $M_A$ exceed unity and the force-free assumption is not valid.
Therefore plasma and flow effects have to be taken into account, which is
done here in the framework of stationary MHD.

\subsection{Evaluation of the Initial State and Optimization}
\begin{figure}
\includegraphics[width=1.0 \textwidth]{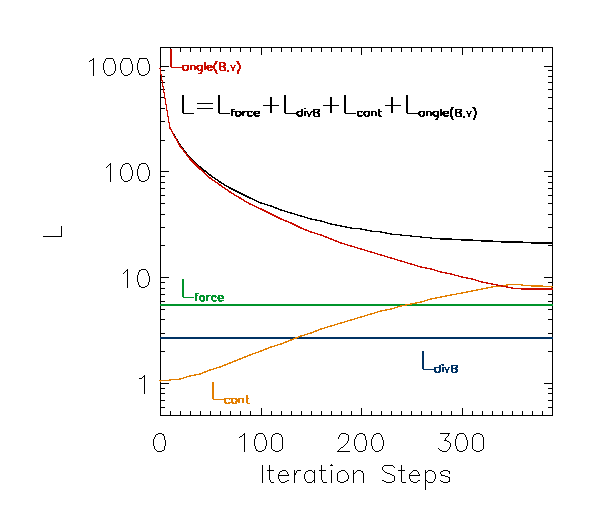}
\caption{In the initial state the force-free
magnetic field and plasma properties are decoupled.
During the iteration B and V become parallel.
In the final state we have found a solution of the stationary
compressible MHD equations with field aligned plasma flow.
The residual discretization errors are similar
to those of nonlinear
force-free computations.}
\label{fig3}
\end{figure}
\begin{table}
\begin{tabular}{l|rrrrrr}
 &$L$&$L_{\rm force}$ & $L_{\rm divB}$ & $L_{\rm cont}$&$L_{\rm angle(B, v)}$
 & Angle \\ \hline
NLFFF, Parker & $973.6$ &  $5.6$ & $ 2.7$ & $1.1$ & $964.2$ & $31.1^{\circ}$ \\
Stationary MHD & $24.6$ & $5.6$ & $ 2.7$ & $8.3$ & $8.0$ & $2.0^{\circ}$ \\
\end{tabular}
\caption{Value of the different terms of the functional $L$
(see Equations \ref{defL}-\ref{L_angle}) for the initial state and the
final stationary MHD-equilibrium. The evolution of these quantities
during the iteration is shown in Figure \ref{fig3}. We monitor
also the weighted angle between magnetic field and plasma flow
as defined in Equation \ref{defangle}.}
\label{table1}
\end{table}
In the initial state the force-free Equations
\ref{nlfff1}-\ref{nlfff2} and the hydrodynamic Equations
\ref{parker1}-\ref{parker3} are fulfilled separately.
Table \ref{table1}
shows in the line 'NLFFF, Parker' the
discretisation errors of the terms
$L_{\rm force}$, $L_{\rm divB}$ and $L_{\rm cont}$
in the initial state.
The term $L_{\rm angle(B, v)}$ is not a discretisation error,
it indicates how well the magnetic field and  the plasma flow are aligned.
For a perfect equilibrium with field aligned flow all terms would be zero.
While this initial state contains a force-free magnetic
field configuration and a spherically
symmetric hydrodynamic solar wind, the
magnetic field
and plasma flow are not aligned.
The averaged weighted
angle between ${\bf v}$ and ${\bf B}$ is $31.1^{\circ}$ and the corresponding
term of the functional $L_{\rm angle(B, v)}$ is more than two orders of
magnitude larger than the
discretisation errors
in final state after relaxation.

We minimize the functional $L$ iteratively with a steepest descent method,
which has been successfully used in
the global nonlinear force-free and
magneto-hydro-static optimization
codes \citep[see][for details]{2007SoPh..240..227W,2007A&A...475..701W}.
Figure \ref{fig3} shows the evolution of $L$ in black and
 its individual
terms: $L_{\rm force}$ in green,$L_{\rm divB}$ in blue,
$L_{\rm cont}$ in orange and $L_{\rm angle(B, v)}$ in red.
While the $L_{\rm force}$ and $L_{\rm divB}$ term stay approximately
constant, the term $L_{\rm angle(B, v)}$  decreases rapidly until
it is of the order of the discretisation errors of the initial equilibrium,
which corresponds to an average weighted angle between flow and magnetic
field of $2^{\circ}$. An exception is the term $L_{\rm cont}$ which increases
somewhat. To understand the evolution of the $L_{\rm cont}$ term
 we  must consider that while the initial
force-free magnetic field is already complex, the plasma flow is smooth
and strictly radial in the initial state, which results in a low initial value
of $L_{\rm cont}$. This is not the case in the final equilibrium, when the
flow becomes field aligned and thereby more complex.
In the final state all terms of $L$ are roughly of the same order and
of the level of the discretisation error of the initial NLFFF-solution
(see also Table \ref{table1}, line 'Stationary MHD').
It can therefore
be considered that a solution of the stationary MHD equations
with field aligned plasma flow has been reached to a good approximation.
\subsection{Brief Analysis of the Final Stationary MHD Equilibrium}
\begin{figure}[h]
\includegraphics[width=0.9 \textwidth]{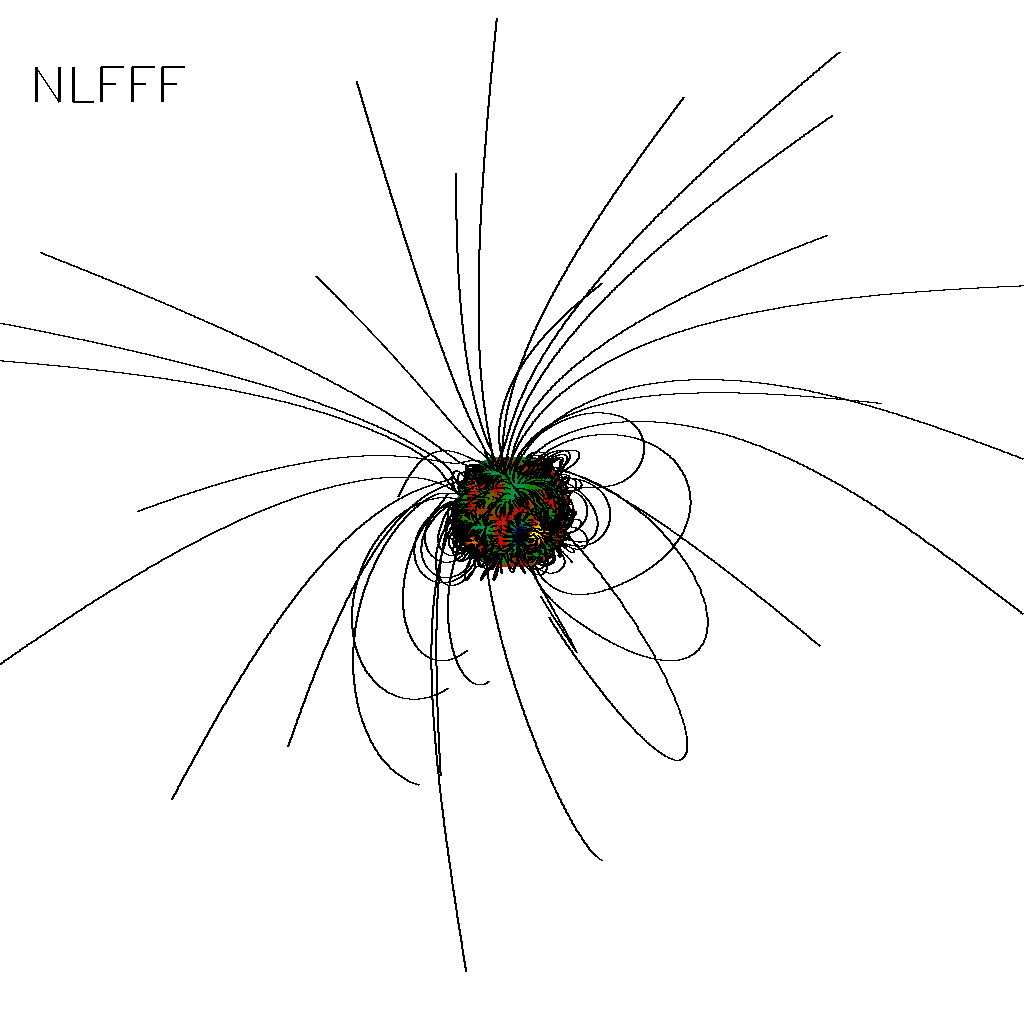}
\includegraphics[width=0.9 \textwidth]{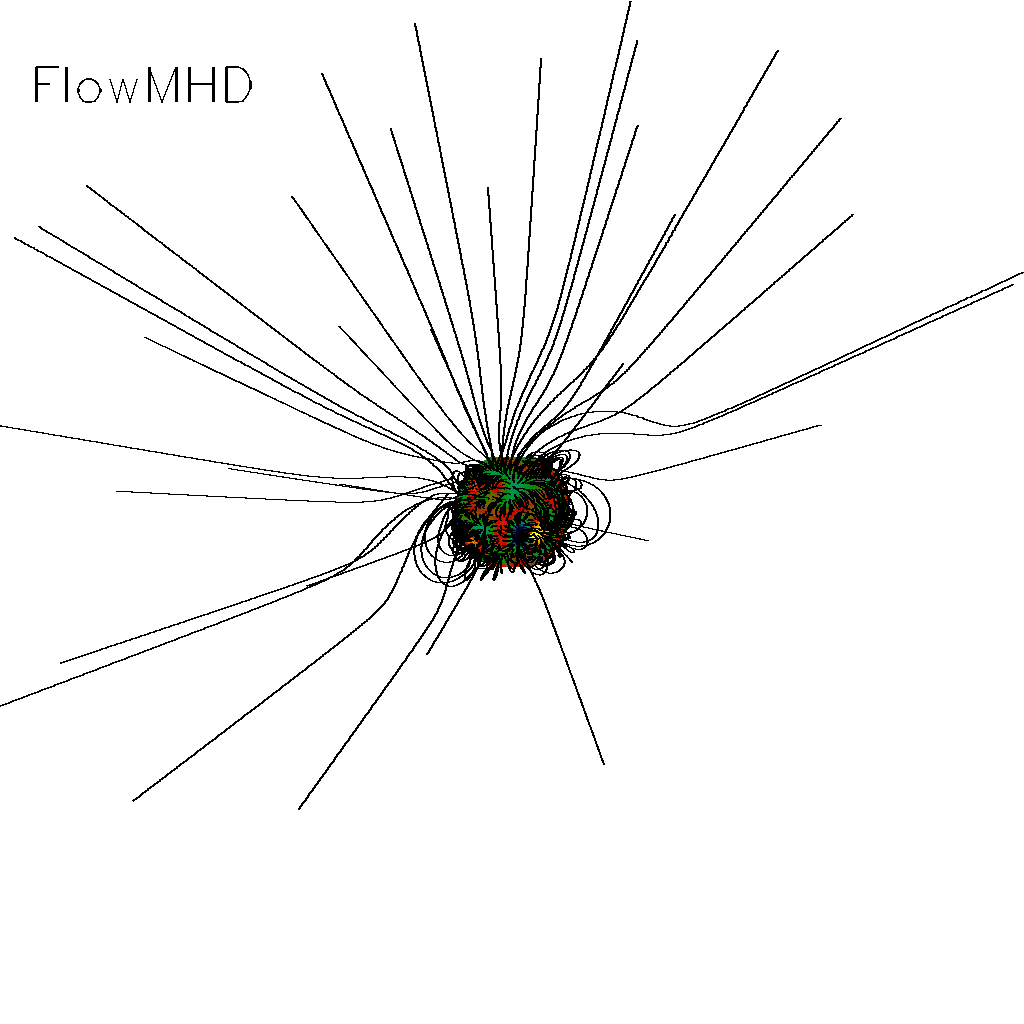}
\caption{Comparison of magnetic field lines for the initial force-free
field [NLFFF] and the final stationary MHD equilibrium [FlowMHD].
Solar wind in  stationary MHD stretches and opens magnetic field lines.
}
\label{fig4}
\end{figure}

\begin{figure}[h]
\mbox{
\includegraphics[width=0.45\textwidth]{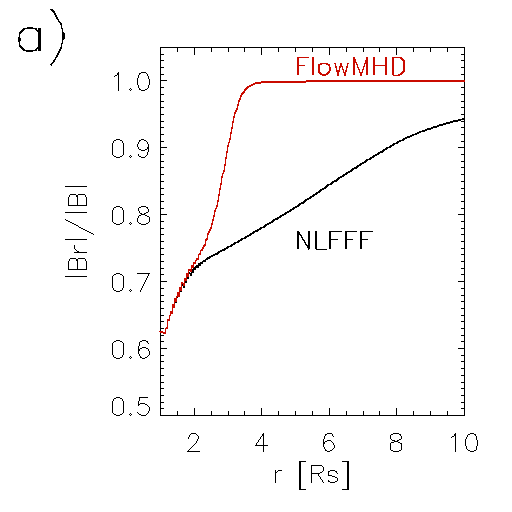}
\includegraphics[width=0.45 \textwidth]{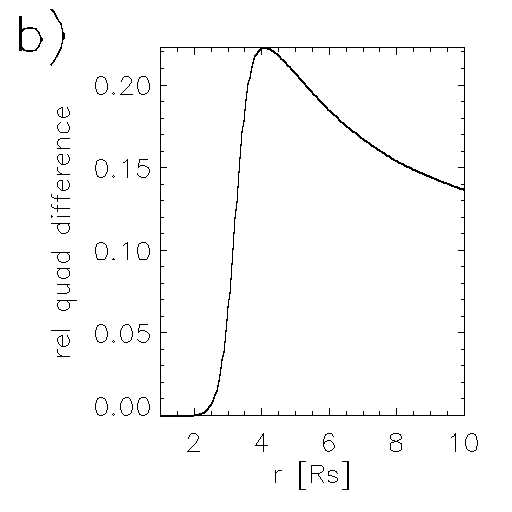}}
\caption{Panel a) Compared with the initial NLFFF the stationary MHD
field becomes more radial
from about 2 solar radii on.
Panel b) Comparison of (horizontal averaged) magnetic field. NLFFF
and stationary MHD are almost
identical below 2 solar radii, where
$M_A \ll 1$ and flow effects are not important.
}
\label{fig5}
\end{figure}

\begin{figure}[h]
\includegraphics[width=12cm, height=6cm, bb = 20 0 980 500, clip]{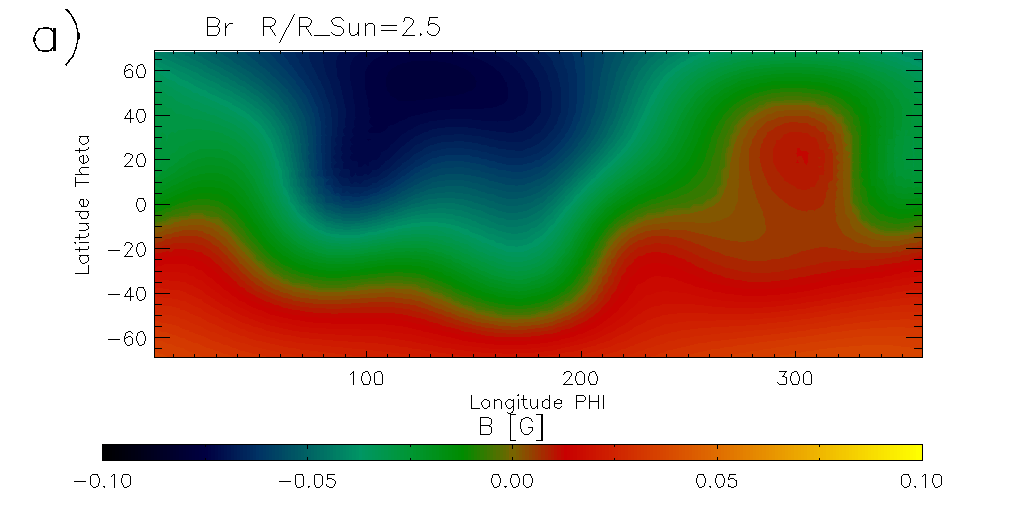}
\includegraphics[width=12cm, height=6cm, bb = 20 0 980 500, clip]{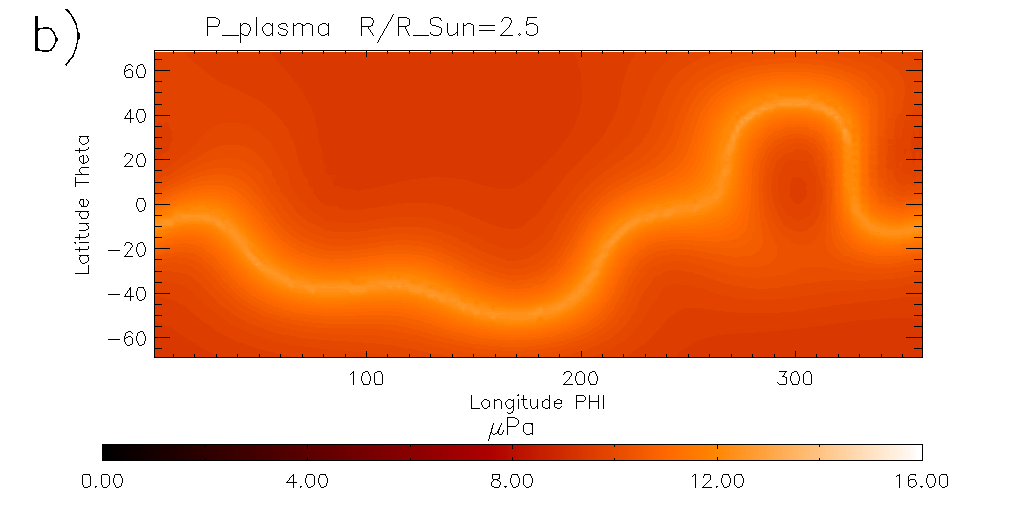}
\includegraphics[width=12cm, height=6cm, bb = 20 0 980 500, clip]{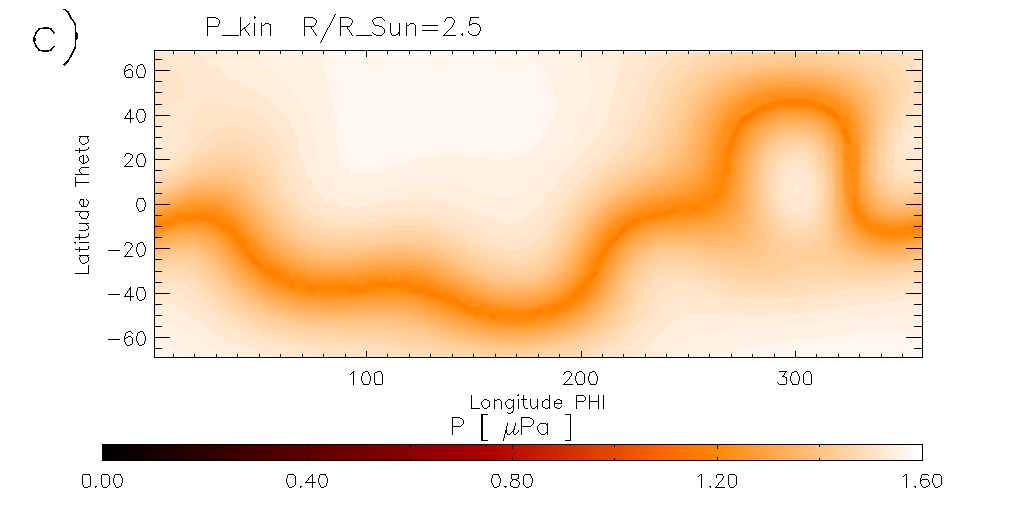}
\caption{Along the polarity inversion line of the magnetic field (panel a)
the plasma pressure (panel b) becomes enhanced and the
kinematic pressure (panel c) decreases.
}
\label{fig6}
\end{figure}

Figure \ref{fig4} shows in the top panel `NLFFF' selected field lines
for the initial force-free equilibrium. In the bottom panel `FlowMHD'
field lines (same starting points) of the final stationary MHD equilibrium
are shown. It seems that low lying field lines are hardly
affected, whereas further
away from the sun the field lines become considerably
more radial in the stationary MHD equilibrium. Physically this can be
understood in the sense that the solar wind flow stretches and
opens the field lines.
In Figure \ref{fig5} we investigate the differences more quantitatively.
Panel a) shows how radial the magnetic field is as a function of the radius,
where $|B_r|$ and $|{\bf B}|$ have been averaged over the whole sphere.
The black line `NLFFF' shows the initial force-free field and the red
line `FlowMHD' the stationary MHD solution.
A value of unity means that
the field is fully radial while a value of zero means the field is
horizontal.
The plot corroborates that the magnetic field becomes more radial
on average with increasing distance from the Sun.
At low heights both solutions
agree, but above about $2 R_s$ the stationary MHD solution becomes
significantly more
radial then the initial field, as was already qualitatively seen in the
field line plots in Figure \ref{fig4}.
Figure \ref{fig5} panel b) shows the relative quadratic difference
(quadratic difference of the magnetic field vectors averaged over
the entire sphere at each radial distance and divided by the averaged
magnetic field strength). This plot confirms
that below about $2 R_s$ the initial force-free and final stationary MHD
equilibria are almost identical and deviate higher up in the corona.

Figure \ref{fig6} shows some quantities at
a height of $r=2.5 R_s$,
which is usually the distance the source surface in located in PFSS-models.
Panel a) shows the radial magnetic field component $B_r$. Between the
positive (red) and negative (green) magnetic field areas is the polarity
inversion line. Around this line the plasma pressure increases (panel b)
and the kinematic pressure
$p_{\rm kin}=\rho v^2/2$
of the solar wind decreases (panel c).
A reduced kinematic pressure around the polarity inversion line
happens naturally, because here the magnetic field and consequently the
aligned plasma flow are not dominated by the radial component.

\subsection{Influence of Initial Conditions}
\label{sec_rev}
\begin{table}[h]
\begin{tabular}{lllll|rrr}
Area & Initial B & Initial T & $c_s$ & $r_c$ & $10^{-4}$&  $10^{-3}$ &$10^{-2}$ \\ \hline
Global & NLFFF & $3$ MK & 157 km/s & $3.84 R_s$ & $1.80 R_s$&$2.11 R_s$&$2.54 R_s$ \\
AR     & NLFFF & $3$ MK & 157 km/s & $3.84 R_s$ & $1.87 R_s$&$2.22 R_s$&$2.57 R_s$ \\
QS     & NLFFF & $3$ MK & 157 km/s & $3.84 R_s$ & $1.77 R_s$&$2.08 R_s$&$2.53 R_s$ \\
\hline
Global & Potential & $3$ MK & 157 km/s & $3.84 R_s$ & $1.84 R_s$&$2.15 R_s$&$2.61 R_s$ \\
AR     & Potential & $3$ MK & 157 km/s & $3.84 R_s$ & $1.91 R_s$&$2.26 R_s$&$2.61 R_s$ \\
QS     & Potential & $3$ MK & 157 km/s & $3.84 R_s$ & $1.80 R_s$&$2.12 R_s$&$2.61 R_s$ \\
\hline
Global & NLFFF & $2$ MK & 129 km/s & $5.77 R_s$ & $2.82 R_s$&$3.20 R_s$&$3.62 R_s$ \\
AR     & NLFFF & $2$ MK & 129 km/s & $5.77 R_s$ & $2.82 R_s$&$3.10 R_s$&$3.44 R_s$ \\
QS     & NLFFF & $2$ MK & 129 km/s & $5.77 R_s$ & $2.96 R_s$&$3.58 R_s$&$4.35 R_s$ \\
\hline
Global & NLFFF & $4$ MK & 182 km/s & $2.88 R_s$ & $1.45 R_s$&$1.62 R_s$&$1.91 R_s$ \\
AR     & NLFFF & $4$ MK & 182 km/s & $2.88 R_s$ & $1.45 R_s$&$1.63 R_s$&$1.98 R_s$ \\
QS     & NLFFF & $4$ MK & 182 km/s & $2.88 R_s$ & $1.42 R_s$&$1.59 R_s$&$1.87 R_s$ \\
\end{tabular}
\caption{Comparison of initial and final magnetic
field. The first four columns contain information
on the initial state and the investigated area.
Column one names the investigated area,
either the entire
sphere `Global', or an active region `AR', or a quiet Sun area `QS'
(see the white and black boxes in Figure \ref{fig1}.
Column two
contains the initial magnetic field model, either a potential field
`Potential' or a nonlinear force-free field `NLFFF'. The third, fourth
and fifth column contain the temperature on the initial hydrodynamic
equilibrium, the sound velocity $c_s$ and the related critical radius
$r_c$ where the solar
wind passes the sound velocity. The remaining columns six, seven and eight
contain the results. Computed is the relative quadratic differences of
the initial and final magnetic field as a function of the radius and
subsequently it is checked at which radial distance from the sun the
difference exceeds $10^{-4}$,  $10^{-3}$ and $10^{-2}$, respectively.}
\label{table2}
\end{table}
An important question is to which extent simpler force-free extrapolations
are justified in some areas. From the equilibrium investigated so
far, it seems that non-force-free effects become important above about $2 R_s$.
In the following we want to quantify this further and investigate
how the radii where the transition from force-free to non-force-free
occur depend on the initial conditions. We also investigate if
the results based on global averaging (marked `Global' in Table \ref{table2})
are different from a selected active region (marked `AR' in
Table \ref{table2} and with a white box in Figure \ref{fig1} top panel) and
a selected quiet sun area
(marked `QS' in Table \ref{table2} and with a black box in Figure
\ref{fig1}). We investigate the effect of the initial magnetic field
(either a nonlinear force-free field (marked with NLFFF) or
a potential field (marked Potential). We also investigate the effect
of the initial hydrodynamic Parker solar wind equilibrium by varying
the temperature.
Similar as in Figure \ref{fig5} panel b) we
compute the relative quadratic difference of the initial magnetic field
and the final stationary MHD-equilibrium. In Table \ref{table2}
we check at which radii the relative quadratic differences exceed
thresholds of $10^{-4}$,  $10^{-3}$,  and $10^{-2}$, respectively.

The areas with $r$ below the radius defined by a relative
quadratic difference of  $10^{-4}$ can be considered as force-free.
For areas with $r$ larger than the radius  defined by a relative
quadratic difference of  $10^{-2}$ the effects of plasma flow and
plasma forces are important. Between the two radii the transition
between force-freeness and stationary MHD take place.
For the equilibrium investigated earlier (first row in Table
\ref{table2}) this transition area is between $1.80 R_s$ and
$2.54 R_s$ for the global case. Restricting the computations on
a selected active region or quiet sun area hardly changes anything.
The corresponding radii change only by less than $0.1 R_s$
(upward for the active region and downward for the quiet sun).
The influence of using a potential field as initial state instead of
a force-free one is very small, too and leads to a very small
(well below $0.1 R_s$) shift upwards. Changing the initial
solar wind equilibrium  has a significantly stronger effect.
For a lower temperature the sound velocity $c_s$ decreases and
the  critical radius $r_c$ where the flow transits the sound velocity
increases. The opposite happens for an increased temperature.
This results in a higher sound velocity and the radius $r_c$, where
the wind passes the sound velocity is lower. Consequently one has significant
faster flows at lower heights. It is found that the
plasma flow starts having a significant effect already at a radial solar
distance of approximately $r_c/2$.
\section{Conclusions and Outlook}
\label{conclusions}
In this paper we developed an optimization principle for the
computation of stationary MHD equilibria, using synoptic vector
magnetograms as boundary condition. As a first step we used a rather simple
spherically symmetric and isothermal solar wind model as initial condition
for the plasma and flow variables. We found that the newly developed code
converged towards a stationary MHD equilibrium because the residual
errors of the equilibrium are comparable with discretisation errors of
nonlinear force-free solution. Below about $2R_s$ the MHD-solution is very
similar to a nonlinear force-free field. The reason is that because of the low
plasma $\beta$ and small Alfv\'en Mach number $M_A$ in these regions,
plasma and flow effects can be neglected. Higher up in the corona the
solar wind flow stretches out and opens up the magnetic field line.
Such an effect of the solar wind is considered already in PFSS-model, but
with an artificial source surface, whereas in stationary
MHD the influence of the wind flow on the magnetic field is computed
self-consistently. While the choice of the initial magnetic field
configuration hardly influences this result, the profile of the outflow
velocity is important. Fast flows in lower heights do obviously require that
the effects of these flows have to be taken into account.

In the application of the method there is certainly room for improvement.
Naturally one could relax the isothermal
condition used here just for reasons of
simplicity and use a more involved energy equation.
In principle it is also possible to take measurements into account for
the solar wind profile, which does not necessarily have
to be spherically symmetric
in the initial state. Including the observations of coronal loops to
constrain the magnetic field as done in the force-free models
of e.g.
\cite{2013SoPh..287..323A,2017ApJ...837...10C} can
in principle also be
incorporated in stationary MHD models. It is a useful feature of
optimization principles
that additional constraints are
straightforward
to implement by additional terms in the functional $L$.

\clearpage
\begin{acks}
TW acknowledges financial support by DLR-grant 50 OC 1701
and DFG-grant WI 3211/5-1.
TN acknowledges financial support by the UK's
Science and Technology Facilities Council (STFC)
via Consolidated Grant ST/S000402/1.
The Astronomical Institute of the Czech Academy of Sciences is supported
by the project RVO:67985815.
IC acknowledges funding by DFG-grant WI 3211/5-1.
We acknowledge stimulating discussions during
two ISSI Team workshops in Bern
led by Anthony Yeates
on 'Global Non-Potential Magnetic Models of the Solar Corona'.
The used  SDO/HMI synoptic vector magnetograms are
courtesy of NASA/SDO and the AIA, EVE, and HMI science teams.
$ $\\\\
{\bf Disclosure of potential conflict of interest:}
The authors declare that they have no conflicts of interest.
\end{acks}
%



\begin{thebibliography}{35}
\ifx\bisbn     \undefined \def\bisbn  #1{ISBN #1}\fi
\ifx\binits    \undefined \def\binits#1{#1}\fi
\ifx\bauthor   \undefined \def\bauthor#1{#1}\fi
\ifx\batitle   \undefined \def\batitle#1{#1}\fi
\ifx\bjtitle   \undefined \def\bjtitle#1{\textit{#1}}\fi
\ifx\bvolume   \undefined \def\bvolume#1{\textbf{#1}}\fi
\ifx\byear     \undefined \def\byear#1{#1}\fi
\ifx\bissue    \undefined \def\bissue#1{#1}\fi
\ifx\bfpage    \undefined \def\bfpage#1{#1}\fi
\ifx\blpage    \undefined \def\blpage #1{#1}\fi
\ifx\burl      \undefined \def\burl#1{\textsf{#1}}\fi
\ifx\href      \undefined \def\href#1#2{\textsf{#2}}\fi
\ifx\betal     \undefined \def\betal{\textit{et al.}}\fi
\ifx\bctitle   \undefined \def\bctitle#1{#1}\fi
\ifx\beditor   \undefined \def\beditor#1{#1}\fi
\ifx\bbtitle   \undefined \def\bbtitle#1{\textit{#1}}\fi
\ifx\bedition  \undefined \def\bedition#1{#1}\fi
\ifx\bseriesno \undefined \def\bseriesno#1{\textbf{#1}}\fi
\ifx\blocation \undefined \def\blocation#1{#1}\fi
\ifx\bsertitle \undefined \def\bsertitle#1{\textit{#1}}\fi
\ifx\bsnm      \undefined \def\bsnm#1{#1}\fi
\ifx\bsuffix   \undefined \def\bsuffix#1{#1}\fi
\ifx\bparticle \undefined \def\bparticle#1{#1}\fi
\ifx\barticle  \undefined \def\barticle#1{}\fi
\ifx\binstitute  \undefined \def\binstitute#1{#1}\fi
\ifx\bpublisher  \undefined \def\bpublisher#1{#1}\fi
\ifx\doiurl    \undefined \def\doiurl#1{\href{#1}{\textsf{DOI}}}\fi
\makeatletter
\def\safeHref#1#2#3{\in@{http}{#2}\ifin@\href{#2}{#3}\else\href{#1#2}{#3}\fi}
\makeatother
\ifx\adsurl    \undefined
  \def\adsurl#1{\safeHref{https://ui.adsabs.harvard.edu/abs/}{#1}{\textsf{ADS}}}\fi
\ifx\arxivurl  \undefined
  \def\arxivurl#1{\safeHref{http://arxiv.org/abs/}{#1}{\textsf{arXiv}}}\fi
\ifx\botherref \undefined \def\botherref#1{}\fi
\ifx\url       \undefined \def\url#1{\textsf{#1}}\fi
\ifx\bchapter  \undefined \def\bchapter#1{}\fi
\ifx\bbook     \undefined \def\bbook#1{}\fi
\ifx\bcomment  \undefined \def\bcomment#1{#1}\fi
\ifx\oauthor   \undefined \def\oauthor#1{#1}\fi
\ifx\citeauthoryear \undefined\def \citeauthoryear#1{#1}\fi
\def\endbibitem {}
\ifx\bconflocation  \undefined \def\bconflocation#1{#1} \fi

\bibitem[\protect\citeauthoryear{{Amari}
  \textit{et~al.}}{2013}]{2013A&A...553A..43A}
\begin{barticle}
\bauthor{\bsnm{{Amari}}, \binits{T.}},
\bauthor{\bsnm{{Aly}}, \binits{J.-J.}},
\bauthor{\bsnm{{Canou}}, \binits{A.}},
\bauthor{\bsnm{{Mikic}}, \binits{Z.}}:
\byear{2013},
\batitle{{Reconstruction of the solar coronal magnetic field in spherical
  geometry}}.
\bjtitle{\aap}
\bvolume{553},
\bfpage{A43}.
\doiurl{https://doi.org/10.1051/0004-6361/201220787}.
\adsurl{2013A&A...553A..43A}.
\end{barticle}
\endbibitem

\bibitem[\protect\citeauthoryear{{Amari}
  \textit{et~al.}}{2014}]{2014JPhCS.544a2012A}
\begin{bchapter}
\bauthor{\bsnm{{Amari}}, \binits{T.}},
\bauthor{\bsnm{{Aly}}, \binits{J.-J.}},
\bauthor{\bsnm{{Chopin}}, \binits{P.}},
\bauthor{\bsnm{{Canou}}, \binits{A.}},
\bauthor{\bsnm{{Mikic}}, \binits{Z.}}:
\byear{2014},
\bctitle{{Large scale reconstruction of the solar coronal magnetic field}}.
In: \bbtitle{Journal of Physics Conference Series},
\bsertitle{Journal of Physics Conference Series}
\bseriesno{544},
\bfpage{012012}.
\doiurl{https://doi.org/10.1088/1742-6596/544/1/012012}.
\adsurl{2014JPhCS.544a2012A}.
\end{bchapter}
\endbibitem

\bibitem[\protect\citeauthoryear{{Aschwanden}}{2013}]{2013SoPh..287..323A}
\begin{barticle}
\bauthor{\bsnm{{Aschwanden}}, \binits{M.J.}}:
\byear{2013},
\batitle{{A Nonlinear Force-Free Magnetic Field Approximation Suitable for Fast
  Forward-Fitting to Coronal Loops. I. Theory}}.
\bjtitle{\solphys}
\bvolume{287}(\bissue{1-2}),
\bfpage{323}.
\doiurl{https://doi.org/10.1007/s11207-012-0069-7}.
\adsurl{2013SoPh..287..323A}.
\end{barticle}
\endbibitem

\bibitem[\protect\citeauthoryear{{Bogdan} and {Low}}{1986}]{bogdan:etal86}
\begin{barticle}
\bauthor{\bsnm{{Bogdan}}, \binits{T.J.}},
\bauthor{\bsnm{{Low}}, \binits{B.C.}}:
\byear{1986},
\batitle{{The three-dimensional structure of magnetostatic atmospheres. II -
  Modeling the large-scale corona}}.
\bjtitle{\apj}
\bvolume{306},
\bfpage{271}.
\doiurl{https://doi.org/10.1086/164341}.
\adsurl{1986ApJ...306..271B}.
\end{barticle}
\endbibitem

\bibitem[\protect\citeauthoryear{{Chifu}, {Wiegelmann}, and
  {Inhester}}{2017}]{2017ApJ...837...10C}
\begin{barticle}
\bauthor{\bsnm{{Chifu}}, \binits{I.}},
\bauthor{\bsnm{{Wiegelmann}}, \binits{T.}},
\bauthor{\bsnm{{Inhester}}, \binits{B.}}:
\byear{2017},
\batitle{{Nonlinear Force-free Coronal Magnetic Stereoscopy}}.
\bjtitle{\apj}
\bvolume{837}(\bissue{1}),
\bfpage{10}.
\doiurl{https://doi.org/10.3847/1538-4357/aa5b9a}.
\adsurl{2017ApJ...837...10C}.
\end{barticle}
\endbibitem

\bibitem[\protect\citeauthoryear{{Contopoulos}}{2013}]{2013SoPh..282..419C}
\begin{barticle}
\bauthor{\bsnm{{Contopoulos}}, \binits{I.}}:
\byear{2013},
\batitle{{The Force-Free Electrodynamics Method for the Extrapolation of
  Coronal Magnetic Fields from Vector Magnetograms}}.
\bjtitle{\solphys}
\bvolume{282}(\bissue{2}),
\bfpage{419}.
\doiurl{https://doi.org/10.1007/s11207-012-0154-y}.
\adsurl{2013SoPh..282..419C}.
\end{barticle}
\endbibitem

\bibitem[\protect\citeauthoryear{{Contopoulos}, {Kalapotharakos}, and
  {Georgoulis}}{2011}]{2011SoPh..269..351C}
\begin{barticle}
\bauthor{\bsnm{{Contopoulos}}, \binits{I.}},
\bauthor{\bsnm{{Kalapotharakos}}, \binits{C.}},
\bauthor{\bsnm{{Georgoulis}}, \binits{M.K.}}:
\byear{2011},
\batitle{{Nonlinear Force-Free Reconstruction of the Global Solar Magnetic
  Field: Methodology}}.
\bjtitle{\solphys}
\bvolume{269}(\bissue{2}),
\bfpage{351}.
\doiurl{https://doi.org/10.1007/s11207-011-9713-x}.
\adsurl{2011SoPh..269..351C}.
\end{barticle}
\endbibitem

\bibitem[\protect\citeauthoryear{{Cranmer}}{2004}]{2004AmJPh..72.1397C}
\begin{barticle}
\bauthor{\bsnm{{Cranmer}}, \binits{S.R.}}:
\byear{2004},
\batitle{{New views of the solar wind with the Lambert W function}}.
\bjtitle{American Journal of Physics}
\bvolume{72}(\bissue{11}),
\bfpage{1397}.
\doiurl{https://doi.org/10.1119/1.1775242}.
\adsurl{2004AmJPh..72.1397C}.
\end{barticle}
\endbibitem

\bibitem[\protect\citeauthoryear{{Cranmer}
  \textit{et~al.}}{2015}]{2015RSPTA.37340148C}
\begin{barticle}
\bauthor{\bsnm{{Cranmer}}, \binits{S.R.}},
\bauthor{\bsnm{{Asgari-Targhi}}, \binits{M.}},
\bauthor{\bsnm{{Miralles}}, \binits{M.P.}},
\bauthor{\bsnm{{Raymond}}, \binits{J.C.}},
\bauthor{\bsnm{{Strachan}}, \binits{L.}},
\bauthor{\bsnm{{Tian}}, \binits{H.}},
\bauthor{\bsnm{{Woolsey}}, \binits{L.N.}}:
\byear{2015},
\batitle{{The role of turbulence in coronal heating and solar wind expansion}}.
\bjtitle{Philosophical Transactions of the Royal Society of London Series A}
\bvolume{373}(\bissue{2041}),
\bfpage{20140148}.
\doiurl{https://doi.org/10.1098/rsta.2014.0148}.
\adsurl{2015RSPTA.37340148C}.
\end{barticle}
\endbibitem

\bibitem[\protect\citeauthoryear{{De Rosa}
  \textit{et~al.}}{2009}]{2009ApJ...696.1780D}
\begin{barticle}
\bauthor{\bsnm{{De Rosa}}, \binits{M.L.}},
\bauthor{\bsnm{{Schrijver}}, \binits{C.J.}},
\bauthor{\bsnm{{Barnes}}, \binits{G.}},
\bauthor{\bsnm{{Leka}}, \binits{K.D.}},
\bauthor{\bsnm{{Lites}}, \binits{B.W.}},
\bauthor{\bsnm{{Aschwanden}}, \binits{M.J.}},
\bauthor{\bsnm{{Amari}}, \binits{T.}},
\bauthor{\bsnm{{Canou}}, \binits{A.}},
\bauthor{\bsnm{{McTiernan}}, \binits{J.M.}},
\bauthor{\bsnm{{R{\'e}gnier}}, \binits{S.}},
\bauthor{\bsnm{{Thalmann}}, \binits{J.K.}},
\bauthor{\bsnm{{Valori}}, \binits{G.}},
\bauthor{\bsnm{{Wheatland}}, \binits{M.S.}},
\bauthor{\bsnm{{Wiegelmann}}, \binits{T.}},
\bauthor{\bsnm{{Cheung}}, \binits{M.C.M.}},
\bauthor{\bsnm{{Conlon}}, \binits{P.A.}},
\bauthor{\bsnm{{Fuhrmann}}, \binits{M.}},
\bauthor{\bsnm{{Inhester}}, \binits{B.}},
\bauthor{\bsnm{{Tadesse}}, \binits{T.}}:
\byear{2009},
\batitle{{A Critical Assessment of Nonlinear Force-Free Field Modeling of the
  Solar Corona for Active Region 10953}}.
\bjtitle{\apj}
\bvolume{696}(\bissue{2}),
\bfpage{1780}.
\doiurl{https://doi.org/10.1088/0004-637X/696/2/1780}.
\adsurl{2009ApJ...696.1780D}.
\end{barticle}
\endbibitem

\bibitem[\protect\citeauthoryear{{DeRosa}
  \textit{et~al.}}{2015}]{2015ApJ...811..107D}
\begin{barticle}
\bauthor{\bsnm{{DeRosa}}, \binits{M.L.}},
\bauthor{\bsnm{{Wheatland}}, \binits{M.S.}},
\bauthor{\bsnm{{Leka}}, \binits{K.D.}},
\bauthor{\bsnm{{Barnes}}, \binits{G.}},
\bauthor{\bsnm{{Amari}}, \binits{T.}},
\bauthor{\bsnm{{Canou}}, \binits{A.}},
\bauthor{\bsnm{{Gilchrist}}, \binits{S.A.}},
\bauthor{\bsnm{{Thalmann}}, \binits{J.K.}},
\bauthor{\bsnm{{Valori}}, \binits{G.}},
\bauthor{\bsnm{{Wiegelmann}}, \binits{T.}},
\bauthor{\bsnm{{Schrijver}}, \binits{C.J.}},
\bauthor{\bsnm{{Malanushenko}}, \binits{A.}},
\bauthor{\bsnm{{Sun}}, \binits{X.}},
\bauthor{\bsnm{{R{\'e}gnier}}, \binits{S.}}:
\byear{2015},
\batitle{{The Influence of Spatial resolution on Nonlinear Force-free
  Modeling}}.
\bjtitle{\apj}
\bvolume{811}(\bissue{2}),
\bfpage{107}.
\doiurl{https://doi.org/10.1088/0004-637X/811/2/107}.
\adsurl{2015ApJ...811..107D}.
\end{barticle}
\endbibitem

\bibitem[\protect\citeauthoryear{{Feng}}{2020}]{2020Feng_book}
\begin{bbook}
\bauthor{\bsnm{{Feng}}, \binits{X.}}:
\byear{2020},
\bbtitle{{Magnetohydrodynamic Modeling of the Solar Corona and Heliosphere}}.
\doiurl{https://doi.org/10.1007/978-981-13-9081-4}.
\end{bbook}
\endbibitem

\bibitem[\protect\citeauthoryear{{Feng}
  \textit{et~al.}}{2012}]{2012SoPh..279..207F}
\begin{barticle}
\bauthor{\bsnm{{Feng}}, \binits{X.}},
\bauthor{\bsnm{{Yang}}, \binits{L.}},
\bauthor{\bsnm{{Xiang}}, \binits{C.}},
\bauthor{\bsnm{{Jiang}}, \binits{C.}},
\bauthor{\bsnm{{Ma}}, \binits{X.}},
\bauthor{\bsnm{{Wu}}, \binits{S.T.}},
\bauthor{\bsnm{{Zhong}}, \binits{D.}},
\bauthor{\bsnm{{Zhou}}, \binits{Y.}}:
\byear{2012},
\batitle{{Validation of the 3D AMR SIP-CESE Solar Wind Model for Four
  Carrington Rotations}}.
\bjtitle{\solphys}
\bvolume{279}(\bissue{1}),
\bfpage{207}.
\doiurl{https://doi.org/10.1007/s11207-012-9969-9}.
\adsurl{2012SoPh..279..207F}.
\end{barticle}
\endbibitem

\bibitem[\protect\citeauthoryear{{Low} and {Lou}}{1990}]{1990ApJ...352..343L}
\begin{barticle}
\bauthor{\bsnm{{Low}}, \binits{B.C.}},
\bauthor{\bsnm{{Lou}}, \binits{Y.Q.}}:
\byear{1990},
\batitle{{Modeling Solar Force-free Magnetic Fields}}.
\bjtitle{\apj}
\bvolume{352},
\bfpage{343}.
\doiurl{https://doi.org/10.1086/168541}.
\adsurl{1990ApJ...352..343L}.
\end{barticle}
\endbibitem

\bibitem[\protect\citeauthoryear{{Mackay} and {van
  Ballegooijen}}{2006}]{2006ApJ...641..577M}
\begin{barticle}
\bauthor{\bsnm{{Mackay}}, \binits{D.H.}},
\bauthor{\bsnm{{van Ballegooijen}}, \binits{A.A.}}:
\byear{2006},
\batitle{{Models of the Large-Scale Corona. I. Formation, Evolution, and
  Liftoff of Magnetic Flux Ropes}}.
\bjtitle{\apj}
\bvolume{641}(\bissue{1}),
\bfpage{577}.
\doiurl{https://doi.org/10.1086/500425}.
\adsurl{2006ApJ...641..577M}.
\end{barticle}
\endbibitem

\bibitem[\protect\citeauthoryear{{Mackay} and
  {Yeates}}{2012}]{2012LRSP....9....6M}
\begin{barticle}
\bauthor{\bsnm{{Mackay}}, \binits{D.H.}},
\bauthor{\bsnm{{Yeates}}, \binits{A.R.}}:
\byear{2012},
\batitle{{The Sun's Global Photospheric and Coronal Magnetic Fields:
  Observations and Models}}.
\bjtitle{Living Reviews in Solar Physics}
\bvolume{9}(\bissue{1}),
\bfpage{6}.
\doiurl{https://doi.org/10.12942/lrsp-2012-6}.
\adsurl{2012LRSP....9....6M}.
\end{barticle}
\endbibitem

\bibitem[\protect\citeauthoryear{{Metcalf}
  \textit{et~al.}}{2008}]{2008SoPh..247..269M}
\begin{barticle}
\bauthor{\bsnm{{Metcalf}}, \binits{T.R.}},
\bauthor{\bsnm{{De Rosa}}, \binits{M.L.}},
\bauthor{\bsnm{{Schrijver}}, \binits{C.J.}},
\bauthor{\bsnm{{Barnes}}, \binits{G.}},
\bauthor{\bsnm{{van Ballegooijen}}, \binits{A.A.}},
\bauthor{\bsnm{{Wiegelmann}}, \binits{T.}},
\bauthor{\bsnm{{Wheatland}}, \binits{M.S.}},
\bauthor{\bsnm{{Valori}}, \binits{G.}},
\bauthor{\bsnm{{McTtiernan}}, \binits{J.M.}}:
\byear{2008},
\batitle{{Nonlinear Force-Free Modeling of Coronal Magnetic Fields. II.
  Modeling a Filament Arcade and Simulated Chromospheric and Photospheric
  Vector Fields}}.
\bjtitle{\solphys}
\bvolume{247}(\bissue{2}),
\bfpage{269}.
\doiurl{https://doi.org/10.1007/s11207-007-9110-7}.
\adsurl{2008SoPh..247..269M}.
\end{barticle}
\endbibitem

\bibitem[\protect\citeauthoryear{{Mikic} and
  {Linker}}{1994}]{1994ApJ...430..898M}
\begin{barticle}
\bauthor{\bsnm{{Mikic}}, \binits{Z.}},
\bauthor{\bsnm{{Linker}}, \binits{J.A.}}:
\byear{1994},
\batitle{{Disruption of Coronal Magnetic Field Arcades}}.
\bjtitle{\apj}
\bvolume{430},
\bfpage{898}.
\doiurl{https://doi.org/10.1086/174460}.
\adsurl{1994ApJ...430..898M}.
\end{barticle}
\endbibitem

\bibitem[\protect\citeauthoryear{{Miki{\'c}}
  \textit{et~al.}}{1999}]{1999PhPl....6.2217M}
\begin{barticle}
\bauthor{\bsnm{{Miki{\'c}}}, \binits{Z.}},
\bauthor{\bsnm{{Linker}}, \binits{J.A.}},
\bauthor{\bsnm{{Schnack}}, \binits{D.D.}},
\bauthor{\bsnm{{Lionello}}, \binits{R.}},
\bauthor{\bsnm{{Tarditi}}, \binits{A.}}:
\byear{1999},
\batitle{{Magnetohydrodynamic modeling of the global solar corona}}.
\bjtitle{Physics of Plasmas}
\bvolume{6}(\bissue{5}),
\bfpage{2217}.
\doiurl{https://doi.org/10.1063/1.873474}.
\adsurl{1999PhPl....6.2217M}.
\end{barticle}
\endbibitem

\bibitem[\protect\citeauthoryear{{Neukirch}}{1995}]{neukirch95}
\begin{barticle}
\bauthor{\bsnm{{Neukirch}}, \binits{T.}}:
\byear{1995},
\batitle{{On self-consistent three-dimensional analytic solutions of the
  magnetohydrostatic equations.}}
\bjtitle{\aap}
\bvolume{301},
\bfpage{628}.
\adsurl{1995A\%26A...301..628N}.
\end{barticle}
\endbibitem

\bibitem[\protect\citeauthoryear{{Nickeler}
  \textit{et~al.}}{2014}]{2014A&A...569A..44N}
\begin{barticle}
\bauthor{\bsnm{{Nickeler}}, \binits{D.H.}},
\bauthor{\bsnm{{Karlick{\'y}}}, \binits{M.}},
\bauthor{\bsnm{{Wiegelmann}}, \binits{T.}},
\bauthor{\bsnm{{Kraus}}, \binits{M.}}:
\byear{2014},
\batitle{{Self-consistent stationary MHD shear flows in the solar atmosphere as
  electric field generators}}.
\bjtitle{\aap}
\bvolume{569},
\bfpage{A44}.
\doiurl{https://doi.org/10.1051/0004-6361/201423819}.
\adsurl{2014A&A...569A..44N}.
\end{barticle}
\endbibitem

\bibitem[\protect\citeauthoryear{{Nickeler}
  \textit{et~al.}}{2017}]{2017ApJ...837..104N}
\begin{barticle}
\bauthor{\bsnm{{Nickeler}}, \binits{D.H.}},
\bauthor{\bsnm{{Wiegelmann}}, \binits{T.}},
\bauthor{\bsnm{{Karlick{\'y}}}, \binits{M.}},
\bauthor{\bsnm{{Kraus}}, \binits{M.}}:
\byear{2017},
\batitle{{Electric Current Filamentation Induced by 3D Plasma Flows in the
  Solar Corona}}.
\bjtitle{\apj}
\bvolume{837}(\bissue{2}),
\bfpage{104}.
\doiurl{https://doi.org/10.3847/1538-4357/aa6043}.
\adsurl{2017ApJ...837..104N}.
\end{barticle}
\endbibitem

\bibitem[\protect\citeauthoryear{{Parker}}{1958}]{1958ApJ...128..664P}
\begin{barticle}
\bauthor{\bsnm{{Parker}}, \binits{E.N.}}:
\byear{1958},
\batitle{{Dynamics of the Interplanetary Gas and Magnetic Fields.}}
\bjtitle{\apj}
\bvolume{128},
\bfpage{664}.
\doiurl{https://doi.org/10.1086/146579}.
\adsurl{1958ApJ...128..664P}.
\end{barticle}
\endbibitem

\bibitem[\protect\citeauthoryear{{Schatten}, {Wilcox}, and
  {Ness}}{1969}]{1969SoPh....6..442S}
\begin{barticle}
\bauthor{\bsnm{{Schatten}}, \binits{K.H.}},
\bauthor{\bsnm{{Wilcox}}, \binits{J.M.}},
\bauthor{\bsnm{{Ness}}, \binits{N.F.}}:
\byear{1969},
\batitle{{A model of interplanetary and coronal magnetic fields}}.
\bjtitle{\solphys}
\bvolume{6}(\bissue{3}),
\bfpage{442}.
\doiurl{https://doi.org/10.1007/BF00146478}.
\adsurl{1969SoPh....6..442S}.
\end{barticle}
\endbibitem

\bibitem[\protect\citeauthoryear{{Schrijver}
  \textit{et~al.}}{2006}]{2006SoPh..235..161S}
\begin{barticle}
\bauthor{\bsnm{{Schrijver}}, \binits{C.J.}},
\bauthor{\bsnm{{De Rosa}}, \binits{M.L.}},
\bauthor{\bsnm{{Metcalf}}, \binits{T.R.}},
\bauthor{\bsnm{{Liu}}, \binits{Y.}},
\bauthor{\bsnm{{McTiernan}}, \binits{J.}},
\bauthor{\bsnm{{R{\'e}gnier}}, \binits{S.}},
\bauthor{\bsnm{{Valori}}, \binits{G.}},
\bauthor{\bsnm{{Wheatland}}, \binits{M.S.}},
\bauthor{\bsnm{{Wiegelmann}}, \binits{T.}}:
\byear{2006},
\batitle{{Nonlinear Force-Free Modeling of Coronal Magnetic Fields Part I: A
  Quantitative Comparison of Methods}}.
\bjtitle{\solphys}
\bvolume{235}(\bissue{1-2}),
\bfpage{161}.
\doiurl{https://doi.org/10.1007/s11207-006-0068-7}.
\adsurl{2006SoPh..235..161S}.
\end{barticle}
\endbibitem

\bibitem[\protect\citeauthoryear{{Schrijver}
  \textit{et~al.}}{2008}]{2008ApJ...675.1637S}
\begin{barticle}
\bauthor{\bsnm{{Schrijver}}, \binits{C.J.}},
\bauthor{\bsnm{{DeRosa}}, \binits{M.L.}},
\bauthor{\bsnm{{Metcalf}}, \binits{T.}},
\bauthor{\bsnm{{Barnes}}, \binits{G.}},
\bauthor{\bsnm{{Lites}}, \binits{B.}},
\bauthor{\bsnm{{Tarbell}}, \binits{T.}},
\bauthor{\bsnm{{McTiernan}}, \binits{J.}},
\bauthor{\bsnm{{Valori}}, \binits{G.}},
\bauthor{\bsnm{{Wiegelmann}}, \binits{T.}},
\bauthor{\bsnm{{Wheatland}}, \binits{M.S.}},
\bauthor{\bsnm{{Amari}}, \binits{T.}},
\bauthor{\bsnm{{Aulanier}}, \binits{G.}},
\bauthor{\bsnm{{D{\'e}moulin}}, \binits{P.}},
\bauthor{\bsnm{{Fuhrmann}}, \binits{M.}},
\bauthor{\bsnm{{Kusano}}, \binits{K.}},
\bauthor{\bsnm{{R{\'e}gnier}}, \binits{S.}},
\bauthor{\bsnm{{Thalmann}}, \binits{J.K.}}:
\byear{2008},
\batitle{{Nonlinear Force-free Field Modeling of a Solar Active Region around
  the Time of a Major Flare and Coronal Mass Ejection}}.
\bjtitle{\apj}
\bvolume{675}(\bissue{2}),
\bfpage{1637}.
\doiurl{https://doi.org/10.1086/527413}.
\adsurl{2008ApJ...675.1637S}.
\end{barticle}
\endbibitem

\bibitem[\protect\citeauthoryear{{Tadesse}
  \textit{et~al.}}{2014}]{2014A&A...562A.105T}
\begin{barticle}
\bauthor{\bsnm{{Tadesse}}, \binits{T.}},
\bauthor{\bsnm{{Wiegelmann}}, \binits{T.}},
\bauthor{\bsnm{{Gosain}}, \binits{S.}},
\bauthor{\bsnm{{MacNeice}}, \binits{P.}},
\bauthor{\bsnm{{Pevtsov}}, \binits{A.A.}}:
\byear{2014},
\batitle{{First use of synoptic vector magnetograms for global nonlinear,
  force-free coronal magnetic field models}}.
\bjtitle{\aap}
\bvolume{562},
\bfpage{A105}.
\doiurl{https://doi.org/10.1051/0004-6361/201322418}.
\adsurl{2014A&A...562A.105T}.
\end{barticle}
\endbibitem

\bibitem[\protect\citeauthoryear{{Throumoulopoulos}}{1998}]{1998JPlPh..59..303T}
\begin{barticle}
\bauthor{\bsnm{{Throumoulopoulos}}, \binits{G.N.}}:
\byear{1998},
\batitle{{Nonlinear axisymmetric resistive magnetohydrodynamic equilibria with
  toroidal flow}}.
\bjtitle{Journal of Plasma Physics}
\bvolume{59}(\bissue{2}),
\bfpage{303}.
\doiurl{https://doi.org/10.1017/S0022377897006338}.
\adsurl{1998JPlPh..59..303T}.
\end{barticle}
\endbibitem

\bibitem[\protect\citeauthoryear{{Throumoulopoulos} and
  {Tasso}}{2000}]{2000JPlPh..64..601T}
\begin{barticle}
\bauthor{\bsnm{{Throumoulopoulos}}, \binits{G.N.}},
\bauthor{\bsnm{{Tasso}}, \binits{H.}}:
\byear{2000},
\batitle{{On resistive magnetohydrodynamic equilibria of an axisymmetric
  toroidal plasma with flow}}.
\bjtitle{Journal of Plasma Physics}
\bvolume{64}(\bissue{5}),
\bfpage{601}.
\doiurl{https://doi.org/10.1017/S0022377800008849}.
\adsurl{2000JPlPh..64..601T}.
\end{barticle}
\endbibitem

\bibitem[\protect\citeauthoryear{{Wheatland}, {Sturrock}, and
  {Roumeliotis}}{2000}]{2000ApJ...540.1150W}
\begin{barticle}
\bauthor{\bsnm{{Wheatland}}, \binits{M.S.}},
\bauthor{\bsnm{{Sturrock}}, \binits{P.A.}},
\bauthor{\bsnm{{Roumeliotis}}, \binits{G.}}:
\byear{2000},
\batitle{{An Optimization Approach to Reconstructing Force-free Fields}}.
\bjtitle{\apj}
\bvolume{540}(\bissue{2}),
\bfpage{1150}.
\doiurl{https://doi.org/10.1086/309355}.
\adsurl{2000ApJ...540.1150W}.
\end{barticle}
\endbibitem

\bibitem[\protect\citeauthoryear{{Wiegelmann}}{2007}]{2007SoPh..240..227W}
\begin{barticle}
\bauthor{\bsnm{{Wiegelmann}}, \binits{T.}}:
\byear{2007},
\batitle{{Computing Nonlinear Force-Free Coronal Magnetic Fields in Spherical
  Geometry}}.
\bjtitle{\solphys}
\bvolume{240}(\bissue{2}),
\bfpage{227}.
\doiurl{https://doi.org/10.1007/s11207-006-0266-3}.
\adsurl{2007SoPh..240..227W}.
\end{barticle}
\endbibitem

\bibitem[\protect\citeauthoryear{{Wiegelmann}, {Petrie}, and
  {Riley}}{2017}]{2017SSRv..210..249W}
\begin{barticle}
\bauthor{\bsnm{{Wiegelmann}}, \binits{T.}},
\bauthor{\bsnm{{Petrie}}, \binits{G.J.D.}},
\bauthor{\bsnm{{Riley}}, \binits{P.}}:
\byear{2017},
\batitle{{Coronal Magnetic Field Models}}.
\bjtitle{\ssr}
\bvolume{210}(\bissue{1-4}),
\bfpage{249}.
\doiurl{https://doi.org/10.1007/s11214-015-0178-3}.
\adsurl{2017SSRv..210..249W}.
\end{barticle}
\endbibitem

\bibitem[\protect\citeauthoryear{{Wiegelmann}
  \textit{et~al.}}{2007}]{2007A&A...475..701W}
\begin{barticle}
\bauthor{\bsnm{{Wiegelmann}}, \binits{T.}},
\bauthor{\bsnm{{Neukirch}}, \binits{T.}},
\bauthor{\bsnm{{Ruan}}, \binits{P.}},
\bauthor{\bsnm{{Inhester}}, \binits{B.}}:
\byear{2007},
\batitle{{Optimization approach for the computation of magnetohydrostatic
  coronal equilibria in spherical geometry}}.
\bjtitle{\aap}
\bvolume{475}(\bissue{2}),
\bfpage{701}.
\doiurl{https://doi.org/10.1051/0004-6361:20078244}.
\adsurl{2007A&A...475..701W}.
\end{barticle}
\endbibitem

\bibitem[\protect\citeauthoryear{{Yeates}}{2014}]{2014SoPh..289..631Y}
\begin{barticle}
\bauthor{\bsnm{{Yeates}}, \binits{A.R.}}:
\byear{2014},
\batitle{{Coronal Magnetic Field Evolution from 1996 to 2012: Continuous
  Non-potential Simulations}}.
\bjtitle{\solphys}
\bvolume{289}(\bissue{2}),
\bfpage{631}.
\doiurl{https://doi.org/10.1007/s11207-013-0301-0}.
\adsurl{2014SoPh..289..631Y}.
\end{barticle}
\endbibitem

\bibitem[\protect\citeauthoryear{{Yeates}
  \textit{et~al.}}{2018}]{2018SSRv..214...99Y}
\begin{barticle}
\bauthor{\bsnm{{Yeates}}, \binits{A.R.}},
\bauthor{\bsnm{{Amari}}, \binits{T.}},
\bauthor{\bsnm{{Contopoulos}}, \binits{I.}},
\bauthor{\bsnm{{Feng}}, \binits{X.}},
\bauthor{\bsnm{{Mackay}}, \binits{D.H.}},
\bauthor{\bsnm{{Miki{\'c}}}, \binits{Z.}},
\bauthor{\bsnm{{Wiegelmann}}, \binits{T.}},
\bauthor{\bsnm{{Hutton}}, \binits{J.}},
\bauthor{\bsnm{{Lowder}}, \binits{C.A.}},
\bauthor{\bsnm{{Morgan}}, \binits{H.}},
\bauthor{\bsnm{{Petrie}}, \binits{G.}},
\bauthor{\bsnm{{Rachmeler}}, \binits{L.A.}},
\bauthor{\bsnm{{Upton}}, \binits{L.A.}},
\bauthor{\bsnm{{Canou}}, \binits{A.}},
\bauthor{\bsnm{{Chopin}}, \binits{P.}},
\bauthor{\bsnm{{Downs}}, \binits{C.}},
\bauthor{\bsnm{{Druckm{\"u}ller}}, \binits{M.}},
\bauthor{\bsnm{{Linker}}, \binits{J.A.}},
\bauthor{\bsnm{{Seaton}}, \binits{D.B.}},
\bauthor{\bsnm{{T{\"o}r{\"o}k}}, \binits{T.}}:
\byear{2018},
\batitle{{Global Non-Potential Magnetic Models of the Solar Corona During the
  March 2015 Eclipse}}.
\bjtitle{\ssr}
\bvolume{214}(\bissue{5}),
\bfpage{99}.
\doiurl{https://doi.org/10.1007/s11214-018-0534-1}.
\adsurl{2018SSRv..214...99Y}.
\end{barticle}
\endbibitem

\end{thebibliography}


\end{article}

\end{document}